\documentclass[onecolumn]{aastex631}

\newcommand{\cf}{{\ifmmode{C_{\rm f}}\else{$C_{\rm f}$}\fi}}
\newcommand{\zem}{{\ifmmode{z_{\rm em}}\else{$z_{\rm em}$}\fi}}
\newcommand{\zabs}{{\ifmmode{z_{\rm abs}}\else{$z_{\rm abs}$}\fi}}
\newcommand{\kms}{{\ifmmode{{\rm km~s}^{-1}}\else{km~s$^{-1}$}\fi}}
\newcommand{\vej}{{\ifmmode{v_{\rm ej}}\else{$v_{\rm ej}$}\fi}}
\newcommand{\vrot}{{\ifmmode{v_{\rm rot}}\else{$v_{\rm rot}$}\fi}}
\newcommand{\cm}{{\ifmmode{{\rm cm}^{-1}}\else{cm$^{-1}$}\fi}}
\newcommand{\cmm}{{\ifmmode{{\rm cm}^{-2}}\else{cm$^{-2}$}\fi}}
\newcommand{\cmmm}{{\ifmmode{{\rm cm}^{-3}}\else{cm$^{-3}$}\fi}}
\newcommand{\lya}{Ly$\alpha$}
\newcommand{\lyb}{Ly$\beta$}

\newcommand{\logN}{{\ifmmode{{\rm log}N}\else{log$N$}\fi}}

\newcounter{species} 
\def\ion#1#2{\setcounter{species}{#2}#1$\;${\scriptsize\Roman{species}}\relax}

\shorttitle{AGN Feedback of NAL quasars}
\shortauthors{Misawa et al.}

\graphicspath{{./}{figure/}}

\begin{document}

\title{AGN Feedback Efficiency of NAL Quasars}

\correspondingauthor{Toru Misawa}
\email{misawatr@shinshu-u.ac.jp}

\author[0000-0002-5464-9943]{Toru Misawa}
\affil{Center for General Education, Shinshu University, 3-1-1 Asahi,
  Matsumoto, Nagano 390-8621, Japan}

\author[0000-0003-4877-9116]{Jane C. Charlton}
\affiliation{Department of Astronomy and Astrophysics, Penn State
  University, 525 Davey Lab, 251 Pollock Road, University Park, PA
  16802}
  
\author[0000-0002-3719-940X]{Michael Eracleous}
\affiliation{Department of Astronomy and Astrophysics, Penn State
  University, 525 Davey Lab, 251 Pollock Road, University Park, PA
  16802} \affiliation{Institute for Gavitation and the Cosmos, Penn
  State University, University Park, PA 16802}

\begin{abstract}
We consider if outflowing winds that are detected via narrow
absorption lines (NALs) with FWHM of $<$ 500~\kms\ (i.e., NAL
outflows) in quasar spectra contribute to feedback. As our sample, we
choose 11 NAL systems in eight optically luminous quasars from the NAL
survey of \citet{mis07a}, based on the following selection criteria:
i) they exhibit ``partial coverage'' suggesting quasar origin (i.e.,
{\it intrinsic} NALs), ii) they have at least one low-ionization
absorption line (\ion{C}{2} and/or \ion{Si}{2}), and iii) the
\lya\ absorption line is covered by available spectra.  The results
depend critically on this selection method, which has caveats and
uncertainties associated with it, as we discuss in a dedicated section
of the paper.  Using the column density ratio of the excited and
ground states of \ion{C}{2} and \ion{Si}{2}, we place upper limits on
the electron density as $n_{\rm e}$ $<$ 0.2--18~\cmmm\ and lower
limits on their radial distance from the flux source $R$ as greater
than several hundreds of kpc. We also calculate lower limits on the
mass outflow rate and kinetic luminosity of $\log(\dot{M}/{\rm
  M_{\odot}~s}^{-1}) > 79$--(3.1$\times 10^{5})$ and $\log(\dot{E_{\rm
    k}}/{\rm erg~s}^{-1}) > 42.9$--49.8, respectively. Taking the NAL
selection and these results at face value, the inferred feedback
efficiency can be comparable to or even larger than those of broad
absorption line and other outflow classes, and large enough to
generate significant AGN feedback. However, the question of the
connection of quasar-driven outflows to NAL absorbers at large
distances from the central engine remains open and should be addressed
by future theoretical work.
\end{abstract}


\section{Introduction}\label{sec:intro}

Outflowing winds from active galactic nuclei (AGN) deposit energy and
momentum into the interstellar medium (ISM) and circum-galactic medium
and regulate the star formation activity of their host galaxies. This
could contribute significantly to the evolution of the host galaxies
(i.e., AGN feedback; \citealt{sil98,sca04,cho18}), if the total
kinetic luminosity of the outflowing winds is larger than
$\sim$0.5\%\ \citep{hop10} or $\sim$5\%\ \citep{sca04,di05} of the
quasar's Eddington luminosity.

AGN outflowing winds are often found via broad absorption lines (BALs)
with line widths of $\geq$ 2,000~\kms\ in the spectra of about
15--40\%\ of optically selected quasars (e.g.,
\citealt{wey91,kni08,dai08,gib09,all11}). Their ejection velocities
from the quasars (\vej) range from several hundreds up to 30,000
\kms\ or more \citep{wey91,rod20,ham18}.

Powerful outflows of several quasars with BALs (i.e., BAL quasars)
have kinetic luminosities that satisfy the above feedback conditions
(e.g., \citealt{ara13,bor13,xu19,byu22a,cho20,wal22}). Significant
feedback is especially likely for extremely high-velocity outflows
(EHVOs) with outflow velocities of $\sim0.1c\,$--$\,0.2c$ (e.g.,
\citealt{rod20,vie22}), similar to the Ultra Fast Outflows (UFOs) that
are often detected via absorption lines in the X-ray spectra of AGNs
\citep{tom10,gof13}.

In addition to BALs, narrow absorption lines (NALs) with line widths
of $\leq$~500~\kms\ and their intermediate subclass (mini-BALs) with
line widths of 500 -- 2000~\kms\ have been used to study AGN
outflowing winds \citep{mis07a,nes08,ham11,gan13,cul19,deh24}.  A
substantial fraction of NALs found in quasar spectra do not arise in
gas that is physically related to the quasars themselves ({\it
  intervening} NALs, hereafter); they arise in cosmologically
intervening absorbers such as foreground galaxies, the intergalactic
medium, and Milky Way gas.  But some fraction of NALs have been
associated with quasar outflows ({\it intrinsic} NALs, hereafter)
based partly on the following statistical arguments: 1) the detection
rate of NALs varies depending on the physical properties of the
background quasars (e.g., optical and radio luminosity, radio spectral
index, and radio morphology; \citealt{ric99,ric01}) and 2) a
significant fraction of NALs remains after accounting for the
contributions from cosmologically intervening absorbers and absorbers
associated with quasar host galaxies or their cluster environment
\citep{nes08,wes08}. Additional arguments for such an association
include time variability, partial coverage, or line-locking exhibited
by {\it some} NALs \citep{ham97b,lew23}.  However, identifying
specific intrinsic NALs is not straightforward because they do not
display all the traits of other intrinsic absorption systems, such as
BALs and mini-BALs, at the same time.  Some NALs do vary (a fraction
of $\sim$20--25\%; e.g., \citealt{wis04,nar04}) and some display
partial coverage (e.g., \citealt{mis07a}). NALs that do display
partial coverage do not do so in all the transitions from the same
system. Therefore, the selection of samples of intrinsic NALs has some
inherent ambiguity and any sample selected by these criteria may be
contaminated by intervening NALs.  (we return to these issues in
\S\ref{sec:caveats} and discuss how they affect the work presented
here). In view of the properties described just above, any NALs that
may be intrinsic are unlikely to be associated with the same
portions/regions of quasar outflows that produce BALs and mini-BALs
\citep[for example,][associated the NAL gas with different layers of
  the outflow than the BAL gas; alternatively the NAL gas may be at a
  different distance from the central engine than the BAL
  gas]{gan01}. Moreover, the structure of NAL absorbers is likely to
be different from the structure of BAL and mini-BAL absorbers; in fact
NAL absorbers have been suggested to be filaments with an inner,
low-ionization ``core'' surrounded by an outer, tenuous,
higher-ionization layer \citep[e.g.,][]{cul19} in order to explain the
detection of partial coverage in only some transitions from the same
system.

There are two possible interpretations of the fraction of quasars
hosting BALs, NALs, and mini-BALs.  The first is an orientation
scenario, in which BALs are observed if our line of sight (LoS) to the
continuum source passes through the main, dense stream of the
outflowing wind (e.g., \citealt{elv00,gan01,ham12}). In this scenario,
the difference in line width could depend the inclination angle of our
LoS relative to the outflow direction or on the distance of the gas
producing the different absorption lines from the continuum
source. The latter possibility is bolstered by the association of
blueshifted absorption lines with outflowing gas at a wide range of
distances from the continuum source, from parsecs to hundreds of
kilo-parsecs (e.g., \citealt{dek01,cha15,ito20}).  An alternative idea
is the evolution scenario, in which BAL quasars are in an evolutionary
stage (e.g., just after galaxy merging) in which they are obscured by
dust \citep{far07,lip06}. This scenario is supported by the
observation that the BAL quasar fraction, outflow velocity, and line
width of absorption lines at \zabs\ $\geq$ 6 are a few times larger
than those at lower redshift \citep{bis22,bis23}.

The physical parameters of outflowing winds have been studied through
variability of the strengths and profiles of the absorption lines,
since almost all BALs and mini-BALs vary over a few years in the
quasar's rest frame (e.g., \citealt{gib08,mis14,cap12}). The
probability of variation increases with the time interval between
observations, approaching unity for intervals of a few years
\citep{cap13}. When multiple BAL troughs are present, they often tend
to vary in a coordinated fashion (e.g., \citealt{fil13,hem19}). In
extreme cases, BAL profiles appear or disappear (e.g.,
\citealt{fil12,mac17,sam19}).  Possible origins of such variations
include (a) motion of the absorbing gas parcels across our line of
sight and (b) changes in the ionization state of the absorber (e.g.,
\citealt{ham08,mis07b,rog16,hua19,vie22}).

In past studies, the mass flow rate ($\dot{M}$) and kinetic luminosity
($\dot{E_{\rm k}}$) of outflowing winds have been estimated as
follows: 1) the electron number density ($n_{\rm e}$) is determined
using measurements of troughs from excited states (e.g.,
\citealt{ham01,bor12}), or from the variability time scale of BALs and
mini-BALs (e.g., \citealt{ham97c,mis05}), 2) the ionization parameter
($U$) and total Hydrogen column density ($N_{\rm H}$) are determined
by comparing the observed column densities of various ions to
simulated values using photoionization models (e.g.,
\citealt{xu18,mil20,wal22}), 3) the radial distance of the absorber
from the center ($R$) is then obtained from the values of $n_{\rm e}$
and $U$ (e.g., \citealt{nar04,rog16}), and 4) the knowledge of $R$,
$N_{\rm H}$ and the velocity of the outflow ($v_{\rm ej}$) allows us
to determine the mass flow rate ($\dot{M}$) and kinetic luminosity
($\dot{E_{\rm k}}$) of a given outflow (see \S\ref{sec:caveats} and
\ref{sec:analysis} for details).

While the feedback efficiency of BALs and mini-BALs have been studied
in detail, those of NALs have hardly been examined because most NALs
are so stable that we cannot extract information from time-variability
analysis. However, NALs potentially have large values of $\dot{M}$
($\propto$ \vej) and $\dot{E_{\rm k}}$ ($\propto$ $v_{\rm ej}^3$) and
could contribute significantly to AGN feedback since they appear to be
more common ($\sim$50\%) and have larger outflow velocities (up to
$\sim$0.2c) compared to BAL and mini-BALs \citep{mis07a}.

In recent years, $\dot{M}$ and $\dot{E_{\rm k}}$ of AGN outflows have
been evaluated using absorption lines from ground states (i.e.,
resonance lines) and excited states of ions such as \ion{C}{2},
\ion{Si}{2}, \ion{O}{4}, and \ion{Ne}{5} (e.g.,
\citealt{ara18,byu22a,byu24}). This method has two advantages over
time-variability analysis: i) it requires only a single-epoch spectrum
and ii) it can be applied to absorption lines that do not vary, such
as NALs.

In this study, we investigate the feedback efficiency of quasars using
intrinsic NAL absorbers, in contrast to past studies that focused on
BALs and mini-BALs. We select intrinsic NAL candidates using their
partial coverage signature, which is subject to caveats.  We calculate
the physical parameters using excited and ground state absorption
lines instead of time-variability analysis. We describe the sample
selection in \S\ref{sec:sample} and discuss its caveats in
\S\ref{sec:caveats}, respectively. We present the analysis in
\S\ref{sec:analysis} and our results and discussion in
\S\ref{sec:res-dis}. Finally, \S\ref{sec:summary} summarizes our work.
Throughout the paper, we use a cosmology with
$H_{0}$=69.6~\kms~Mpc$^{-1}$, $\Omega_{m}$=0.286, and
$\Omega_{\Lambda}$=0.714 \citep{ben14}.

\section{Sample Selection}\label{sec:sample}

In this work we rely on partial coverage analysis to select quasars
with intrinsic NALs. This selection method involves several caveats,
which we discuss in \S\ref{sec:caveats} but it is the only method that
can yield a sample of quasars with available high-resolution spectra
suited for the analysis that we carry out here. \citet{mis07a}
identified 39 intrinsic NAL candidates based on partial coverage
analysis of \ion{C}{4}, \ion{N}{5}, and \ion{Si}{4} doublets in the
spectra of 20 bright quasars at \zem\ = 2--4.  Since the sample
searched included 37 quasars, at least 50\%\ of quasars host intrinsic
NALs. The detection limits in rest-frame equivalent width were
$EW_{\rm lim}$(\ion{C}{4}) = 0.056~\AA, $EW_{\rm lim}$(\ion{N}{5}) =
0.038~\AA, and $EW_{\rm lim}$(\ion{Si}{4}) = 0.054~\AA.

The quasars were originally selected for a survey aimed at measuring
the Deuterium-to-Hydrogen abundance ratio (D/H) in the \lya\ forest
(e.g., \citealt{bur98a,bur98b}). Therefore, the target selection does
not have a direct bias with respect to the properties of any intrinsic
absorption-line systems, although there could be indirect bias since
the sample contains only optically bright quasars.  The observations
were carried out with Keck/HIRES with a slit width of
1.$^{\!\!\prime\prime}14$ (i.e., FWHM $\sim$8 \kms). The spectra were
extracted with the automated program, {\sc MAKEE}, written by Tom
Barlow.

We focus on 16 intrinsic NAL candidates of the 39 absorption systems
that have at least one low-ionization absorption line from an ion with
ionization potential between 13 and 24~eV (e.g., \ion{O}{1},
\ion{Si}{2}, \ion{Al}{2}, and \ion{C}{2}).  We reject two systems with
large column densities of $\log N_{\rm HI}$ $>$ 19 since there are
several observational indicators suggesting that some
(sub-)DLAs\footnote{Damped Ly$\alpha$ systems}, contain small-sized
and high-density clouds within themselves: e.g., occurrence of
time-variation \citep{hac13} and emergence of absorption lines from
ions in excited states (e.g., \citealt{wol03}).

Of the remaining 14 systems, we choose 11 systems in 8 quasars as our
sample because i) they have \ion{C}{2} and/or \ion{Si}{2} NALs which
are necessary for the excited/resonance line analysis, and ii) the
\lya\ absorption line of the system is covered by past observations
which is necessary for measuring total column densities of the
outflowing winds.

We use the software package {\sc minfit} \citep{chu97,chu03} to fit
absorption lines with Voigt profiles using the absorption redshift
(\zabs), column density ($\log N$ in \cmm), Doppler parameter ($b$ in
\kms), and covering factor (\cf) as free parameters.

The covering factor, \cf, is the fraction of photons from the
background flux source that pass through the foreground absorber along
our line of sight; it is a measure of the dilution of the depths of
absorption lines and needs to be included to get the correct column
density. We evaluate \cf\ as
\begin{equation}
  C_{\rm f} = \frac{\left( 1 - R_{\rm r} \right)^2}{1 + R_{\rm b} -
    2R_{\rm r}},
\end{equation}
where $R_{\rm b}$ and $R_{\rm r}$ are the residual fluxes at the blue
and red members of a doublet in the normalized spectrum (e.g.,
\citealt{wam95,bar97,ham97c}).  If \cf\ $<$ 1, the absorbers are
likely arising in quasar outflowing winds (i.e., intrinsic absorbers)
because cosmologically intervening absorbers like the CGM of
foreground galaxies and the IGM have sizes much larger than the quasar
continuum source. Based on the results of the covering factor
analysis, \citet{mis07a} separated all NALs into three classes:
classes A and B were deemed to include reliable and possible intrinsic
NALs and class C was taken to include intervening or unclassified
NALs. The detailed classification criteria of each class are
summarized in \citet{mis07a}.

The properties of the quasars in our sample are summarized in
Table~\ref{tab:qsos}. Columns (1) and (2) are the quasar name and
emission redshift (\zem). Columns (3) and (4) give the V- and R-band
magnitudes ($m_{\rm V}$ and $m_{\rm R}$, respectively). Columns
(5)--(8) list the radio-loudness parameter ($\mathcal{R}$), the
bolometric luminosity ($L_{\rm bol}$), the mass of the quasar's
supermassive black hole ($M_{\rm BH}$), and the Eddington luminosity
($L_{\rm Edd}$). We calculate the bolometric luminosity by $L_{\rm
  bol} = 4.4\lambda L_{\lambda}$(1450\AA) (c.f., \citealt{ric06}). We
use SDSS spectra to measure the monochromatic luminosity at
1450\AA\ except for HS1946+7658, for which we use the spectrum of
\citet{hag92}. All spectra are corrected for Galactic extinction
\citep{sch11,car89} before measuring the observed flux. We estimate
the BH mass from the FWHM of the \ion{C}{4} emission line, the
\ion{C}{4} emission line blueshift relative to the systemic redshift,
and $\lambda L_{\rm \lambda}$(1350\AA) using the empirical relation of
\citet{coa17}.

Table~\ref{tab:nals} gives the parameters of intrinsic NALs: column
(1) gives the quasar name, column (2) the absorption redshift (\zabs),
column (3) the neutral Hydrogen column density ($\log N_{\rm HI}$),
columns (4)--(6) the ion of the targeted absorption line (\ion{C}{2}
or \ion{Si}{2}), and the corresponding column densities in the ground
and excited states ($\log N_{\rm l}$ and $\log N_{\rm u}$), columns
(7) and (8) the derived limit on the electron density ($n_{\rm e}$)
and radial distance from the center ($R$), and column (9), the
reliability class of the intrinsic NAL.

\begin{deluxetable}{cccccccc}
\tablecaption{Sample Quasars\label{tab:qsos}}
\tablewidth{0pt}
\tablehead{
\colhead{(1)}   &
\colhead{(2)}   &
\colhead{(3)}   &
\colhead{(4)}   &
\colhead{(5)}   &
\colhead{(6)}   &
\colhead{(7)}   &
\colhead{(8)}   \\
\colhead{quasar}            &
\colhead{\zem}              &
\colhead{$m_{\rm V}$$^a$}     &
\colhead{$m_{\rm R}$$^a$}     &
\colhead{$\mathcal{R}$$^b$} &
\colhead{$L_{\rm bol}$$^c$}   &
\colhead{$M_{\rm BH}$$^d$}    &
\colhead{$L_{\rm Edd}$$^e$}   \\
\colhead{}                  &
\colhead{}                  &
\colhead{(mag)}             &
\colhead{(mag)}             &
\colhead{}                  &
\colhead{(erg~s$^{-1}$)}     &
\colhead{($M_{\odot}$)}      &
\colhead{(erg~s$^{-1}$)}           
}
\startdata
Q0805+0441  & 2.88  & 18.16 &       & 3115    & 1.74$\times 10^{47}$ & 1.94$^{+0.25}_{-0.21} \times 10^{9}$  & 2.45$^{+0.32}_{-0.27} \times 10^{47}$ \\
HS1103+6416 & 2.191 & 15.42 &       & $<$0.62 & 1.18$\times 10^{48}$ & 2.36$^{+0.31}_{-0.26} \times 10^{10}$ & 2.97$^{+0.39}_{-0.32} \times 10^{48}$ \\
Q1107+4847  & 3.000 & 16.60 &       & $<$1.95 & 1.24$\times 10^{48}$ & 3.30$^{+0.38}_{-0.32} \times 10^{9}$  & 4.16$^{+0.48}_{-0.41} \times 10^{47}$ \\
Q1330+0108  & 3.510 &       & 18.56 & $<$16.2 & 3.93$\times 10^{47}$ & 1.69$^{+0.22}_{-0.18} \times 10^{9}$  & 2.13$^{+0.27}_{-0.23} \times 10^{47}$ \\
Q1548+0917  & 2.749 & 18.00 &       & $<$6.96 & 4.20$\times 10^{47}$ & 5.96$^{+0.80}_{-0.66} \times 10^{9}$  & 7.51$^{+1.00}_{-0.84} \times 10^{47}$ \\
Q1554+3749  & 2.664 & 18.19 &       & $<$21.9 & 3.26$\times 10^{47}$ & 3.54$^{+0.44}_{-0.37} \times 10^{9}$  & 4.46$^{+0.55}_{-0.46} \times 10^{47}$ \\
HS1700+6416 & 2.722 & 16.12 &       & $<$1.24 & 2.09$\times 10^{48}$ & 2.59$^{+0.34}_{-0.28} \times 10^{10}$ & 3.27$^{+0.42}_{-0.35} \times 10^{48}$ \\
HS1946+7658 & 3.051 & 16.20 &       & $<$1.35 & 2.87$\times 10^{48}$ & 8.93$^{+1.03}_{-0.88} \times 10^{9}$  & 1.12$^{+0.13}_{-0.11} \times 10^{48}$ \\
\enddata
\tablenotetext{a}{Observed V- or R-band magnitude from \citet{mis07a}.}
\tablenotetext{b}{Radio loudness parameter $\mathcal{R} =
  f_{\nu}$(5~GHz)/$f_{\nu}$(4400~\AA) from \citet{mis07a}.}
\tablenotetext{c}{Bolometric luminosity. See discussion in \S\ref{sec:sample}.}
\tablenotetext{d}{Black hole mass. See discussion in \S\ref{sec:sample}.}
\tablenotetext{e}{Eddington luminosity.}
\end{deluxetable}

\begin{deluxetable}{ccccccccc}
\tablecaption{Properties of NAL Outflows \label{tab:nals}}
\tablewidth{0pt}
\tablehead{
\colhead{(1)}   &
\colhead{(2)}   &
\colhead{(3)}   &
\colhead{(4)}   &
\colhead{(5)}   &
\colhead{(6)}   &
\colhead{(7)}   &
\colhead{(8)}   &
\colhead{(9)}   \\
\colhead{quasar}                &
\colhead{\zabs}                 &
\colhead{$\log N_{\rm HI}$$^a$}   &
\colhead{ION$^b$}               &
\colhead{$\log N_{\rm l}$$^c$}    &
\colhead{$\log N_{\rm u}$$^c$}    &
\colhead{$n_{\rm e}$$^d$}         &
\colhead{$R$$^e$}               &
\colhead{class$^f$}             \\
\colhead{}                      &
\colhead{}                      &
\colhead{(cm$^{-2}$)}            &
\colhead{}                      &
\colhead{(cm$^{-2}$)}            &
\colhead{(cm$^{-2}$)}            &
\colhead{(cm$^{-3}$)}            &
\colhead{(kpc)}                 &
\colhead{}                      
}
\startdata
Q0805+0441  & 2.6517 & 15.11 & \ion{C}{2}  & 13.25 & $<$12.90 & $<$14.38 & $>$108  & A \\
HS1103+6416 & 1.8919 & 15.53 & \ion{Si}{2} & 13.45 & $<$11.78 & $<$18.37 & $>$248  & B \\
Q1107+4847  & 2.7243 & 15.50 & \ion{C}{2}  & 12.70 & $<$12.26 & $<$11.09 & $>$328  & A \\
Q1330+0108  & 3.1148 & 14.99 & \ion{C}{2}  & 12.52 & $<$12.07 & $<$10.78 & $>$187  & A \\
Q1548+0917  & 2.6082 & 18.00 & \ion{C}{2}  & 13.47 & $<$12.31 & $<$1.79  & $>$475  & B \\
            & 2.6659 & 15.26 & \ion{C}{2}  & 13.11 & $<$12.46 & $<$6.30  & $>$253  & A \\
            & 2.6998 & 15.70 & \ion{C}{2}  & 12.67 & $<$12.24 & $<$11.41 & $>$188  & A \\
Q1554+3749  & 2.3777 & 15.63 & \ion{C}{2}  & 13.38 & $<$12.33 & $<$2.33  & $>$367  & A \\
HS1700+6416 & 2.4330 & 18.33 & \ion{C}{2}  & 12.88 & $<$12.02 & $<$3.90  & $>$717  & B \\
HS1946+7658 & 2.8928 & 17.11 & \ion{C}{2}  & 12.34 & $<$11.71 & $<$6.64  & $>$644  & A \\
            & 3.0497 & 17.37 & \ion{C}{2}  & 13.44 & $<$11.35 & $<$0.20  & $>$3712 & A \\
\enddata
\tablenotetext{a}{Column density of neutral Hydrogen.}
\tablenotetext{b}{Ion for which the column density, electron density,
  and absorber's radial distance are reported in columns (5) -- (8).}
\tablenotetext{c}{Column densities of the ion in column (4) in the
  ground and excited states.}
\tablenotetext{d}{Electron density of the ion in column (4).}
\tablenotetext{e}{Absorber's radial distance, corresponding to the ion
  listed in column (4).}
\tablenotetext{f}{NAL classification made in \citet{mis07a}.}
\end{deluxetable}

\section{Caveats Associated With the Sample Selection}\label{sec:caveats}

The conclusions of this work depend critically on whether the NALs
that we have selected are indeed intrinsic.  As we noted in
\S\ref{sec:intro}, NALs do not show all the properties that signal the
association of BALs and mini-BALs with an outflowing wind. Therefore,
we highlight here the uncertainties associated with the partial
coverage analysis method that is the basis of our selection.

\begin{enumerate}

\item
It is possible that compact, dense clouds in a low-ionization state
can exist in over-dense regions in DLA and sub-DLA systems. This
possibility is suggested by (a) studies that find variable NALs at
large blueshifts from the quasars toward which they are observed
\citep[see][]{hac13}, (b) NALs from low-ionization, dense clouds that
may show partial coverage in a handful of DLA or sub-DLA
systems\footnote{These systems also suffer from unresolved saturation,
  but photoionization models identify them as extremely dense and cold
  absorbers corresponding to the molecular gas in the Milky Way.}
\citep[see][]{jones10}, and by a variety of models for the ISM of
galaxies that predict compact, low-ionization clouds
\citep[e.g.,][]{pfenn94,wol03,glover07,fujita09}. Such clouds could
have sizes of a few AU and exhibit partial coverage, although our
targets that exhibit partial coverage in higher-ionized species (i.e.,
\ion{C}{4} and \ion{Si}{4}) are likely to have larger sizes.

\item
Intrinsic NALs that are identified based on partial coverage (\cf\ $<$
1) do not show any other evidence connecting them to intrinsic
absorbers (e.g., time-variability and broad line width) as seen in
BALs and mini-BALs. Therefore, we cannot corroborate that they are
intrinsic by an additional and separate test.

\item
Intrinsic NALs have a black (i.e., fully absorbed) \lya\ trough
(\cf\ = 1) while BAL and mini-BAL systems show a non-black
\lya\ trough (\cf\ $<$ 1). Moreover, in most cases only one component
in the intrinsic NAL systems we have selected shows \cf\ $<$ 1 while
the other components are consistent with \cf\ = 1.

\end{enumerate}

To guard against the first caveat, we exclude DLA and sub-DLA systems
from our sample even if they show partial coverage.  The second and
third caveats introduce uncertainty in our selection of intrinsic NALs
since we have to rely only on the partial coverage signature from one
transition. Nonetheless, we do not that, (a) photoionization models
for intervening absorbers that can produce \ion{C}{4} and \ion{Si}{4}
lines typically have kpc-scale sizes (e.g., \citealt{ste16,sam24}),
and (b) intrinsic absorbers may have a dense, compact core that can
produce partial coverage in some transitions and a more extended halo
that can produce full coverage in others \citep[see][]{cul19}.

\section{Analysis}\label{sec:analysis}

We calculate the mass outflow rate ($\dot{M}$), kinetic luminosity
($\dot{E_{\rm k}}$), and eventually the feedback efficiency following
the prescription of \citet{bor12}. First, we need to estimate the
absorber's radial distance from the ionizing continuum source ($R$),
total Hydrogen column density ($N_{\rm H}$), and ejection velocity
(\vej). Of these parameters, only \vej\ is measured directly from the
observed spectra.

We estimate the absorber's distance $R$ based on the definition of the
ionization parameter\footnote{The ionization parameter is defined as
  $U \equiv n_{\gamma}/n_{\rm H}$, where $n_{\gamma}$ is ionizing
  photon density.}

\begin{equation}
  U = \frac{Q({\rm H})}{4 \pi R^2 n_{\rm H} c},
\end{equation}
where $Q(H)$ is the total number of Hydrogen ionizing photons emitted
per unit time, $n_{\rm H}$ is the Hydrogen number density, and $c$ is
the speed of light.  In order to evaluate accurately the ionization
parameter, it is necessary to use photoionization models.  However,
the number of absorption lines detected in these NAL systems is
limited to only three elements (i.e., H, C, and Si) and the covering
factor of single lines (i.e., \lya, \ion{C}{2}, and \ion{Si}{2})
cannot be evaluated in practice, which prevents us from constraining
the photoionization models effectively. Therefore, we instead assume
$\log U$ = $-2.6$ because that is the ionization state where the
average value of the column density ratio\footnote{The column
  densities of \ion{C}{2} and \ion{C}{4} are evaluated by fitting
  Voigt profiles to their absorption troughs. We evaluate a column
  density of \ion{C}{2} using the same covering factor as
  \ion{C}{4}. We also assume \ion{C}{2} and \ion{C}{4} absorbers have
  same ionization parameter in a single-phase, since their absorption
  profiles are very similar with negligible velocity offset from each
  other.} of \ion{C}{2} and \ion{C}{4} in our NAL systems,
$N$(\ion{C}{2})/$N$(\ion{C}{4}) $\sim$ 0.3, is approximately
reproduced \citep{ham97a}\footnote{We also perform photoionization
  models ourselves for an optically thin cloud with solar abundance
  assuming a conventional quasar SED \citep{nar04} and confirm that
  the ionization condition of $\log U$ = $-2.6$ well reproduces the
  column density ratio of $N$(\ion{C}{2})/$N$(\ion{C}{4}) $\sim$
  0.3.}.  We convert the quasar bolometric luminosity $L_{\rm bol}$ to
$Q(H)$ using a conventional quasar SED that is a segmented power law,
$f_{\nu} \propto \nu^{-\alpha}$, where $\alpha$ = 0.4, 1.6, and 0.9 at
$\log \nu$ = 13.5--15.5, 15.5--17.5, and 17.5--19.5~Hz, respectively
\citet{nar04}.

We also need to know the absorber's Hydrogen number density $n_{\rm
  H}$, which is related to the absorber's electron density $n_{\rm e}$
as $n_{\rm e} \sim 1.2 n_{\rm H}$ in highly ionized plasma
\citep{ost06}.  While \citet{nar04} used the variability time scale as
an upper limit on the recombination time to estimate the electron
density, we use the column density ratio of excited and ground states
of \ion{C}{2} and \ion{Si}{2} following previous works (e.g.,
\citealt{ara13,byu22a}) since i) we have spectra from only a single
epoch, and ii) intrinsic NALs are rarely variable \citep{mis14}.

We estimate the total Hydrogen column density (i.e., $N_{\rm H\;I}$ +
$N_{\rm H\;II}$) from the neutral Hydrogen column density ($N_{\rm
  H\;I}$) using a Cloudy photoionization model (version c17.02,
\citealt{fer17}) with an ionization parameter of $\log U$ = $-2.6$ and
the conventional quasar SED as introduced above.

\begin{figure}[ht!]
  \begin{center}
    \includegraphics[width=9cm,angle=270]{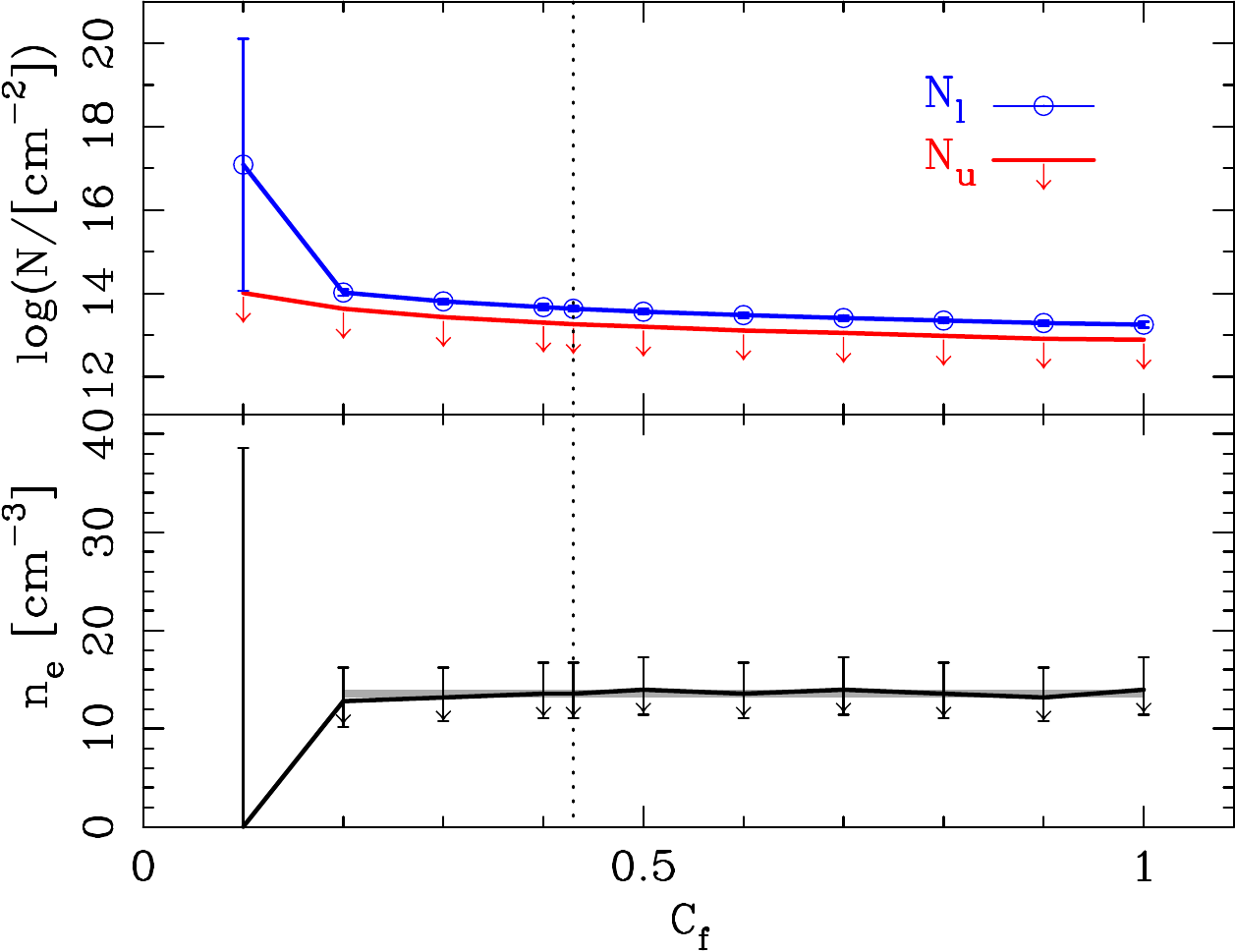}
  \end{center}  
  \caption{Estimated column densities of the ground and excited states
    ($N_{\rm l}$ and $N_{\rm u}$, respectively) of the \ion{C}{2} line
    at \zabs\ = 2.6517 in the spectrum of Q0805+0441 (top panel) and
    their corresponding electron density (bottom panel) as an example
    demonstrating the lack of sensitivity of $n_e$ to \cf. The
    \ion{C}{4} line of the system has covering factor of \cf\ = 0.43
    (marked with a vertical dotted line), but the \ion{C}{2} line
    could have different \cf\ value. Therefore, we estimate column
    densities of \ion{C}{2} and \ion{C}{2}$^*$ as a function of
    covering factor. We place only upper limits on $N$(\ion{C}{2}$^*$)
    and $n_{\rm e}$.  For \cf\ $\geq$ 0.2, an upper limit of $n_{\rm
      e}$ is almost constant ($\leq 13.5\pm 0.4$, considering only
    systematic errors) as shown with the shaded area on the bottom
    panel.\label{fig:ratio}}
\end{figure}

We also use the Voigt profile fitting code {\sc minfit} to measure
column densities as well as Doppler $b$ parameters of ground and
excited states of \ion{C}{2} and \ion{Si}{2}. We assume full coverage
(i.e., \cf\ = 1) because we cannot evaluate \cf\ for single lines.  If
\cf\ $<$ 1, our estimate, which assumes \cf=1, is a lower limit on the
column density.  However, the estimated electron density ratio $N_{\rm
  l} / N_{\rm u}$ and the corresponding electron density would be
insensitive to \cf\ unless the lines are saturated, as demonstrated in
Figure~\ref{fig:ratio}. If the absorption lines require multiple
components to fit their profiles, we only consider the strongest
components for calculating electron densities.  Using the column
density ratio of the strongest components, we place constraints on the
electron density using
\begin{equation}
  n_{\rm e} = n_{\rm cr} \left[ \frac{N_{\rm l}}{N_{\rm u}} \left(
    \frac{g_{\rm u}}{g_{\rm l}} \right) e^{-\Delta E/kT} -1
    \right]^{-1},
\end{equation}
where $n_{\rm cr}$ is the critical density, $N_{\rm l}$ and $N_{\rm
  u}$ are the column densities of the ground and excited states,
$g_{\rm l}$ and $g_{\rm u}$ are their statistical weights ($g_{\rm u}
/ g_{\rm l}$ = 2 for the relevant energy levels of \ion{C}{2} and
\ion{Si}{2}), $\Delta E$ is the energy difference between ground and
excited states, $k$ is the Boltzmann constant, and $T$ is gas
temperature.

Since $\Delta E$ ($1.26 \times 10^{-14}$~erg and $5.69 \times
10^{-14}$~erg for \ion{C}{2} and \ion{Si}{2}\footnote{\tt
  https://www.nist.gov/pml/atomic-spectra-database}) is much smaller
than $kT$ corresponding to the typical temperature of NAL systems
(i.e., $T \sim 10^4$~K), the exponential part of equation~(3) can be
considered as $\sim$1. Using the critical densities of \ion{C}{2} and
\ion{Si}{2} ($n_{\rm cr}$ $\sim$50~\cmmm\ and $\sim$1700~\cmmm\ for
\ion{C}{2} and \ion{Si}{2}; \citealt{tay08a,tay08b}), we can calculate
the electron densities of the NAL systems.

\begin{figure}[ht!]
  \begin{center}
    \includegraphics[width=9cm,angle=270]{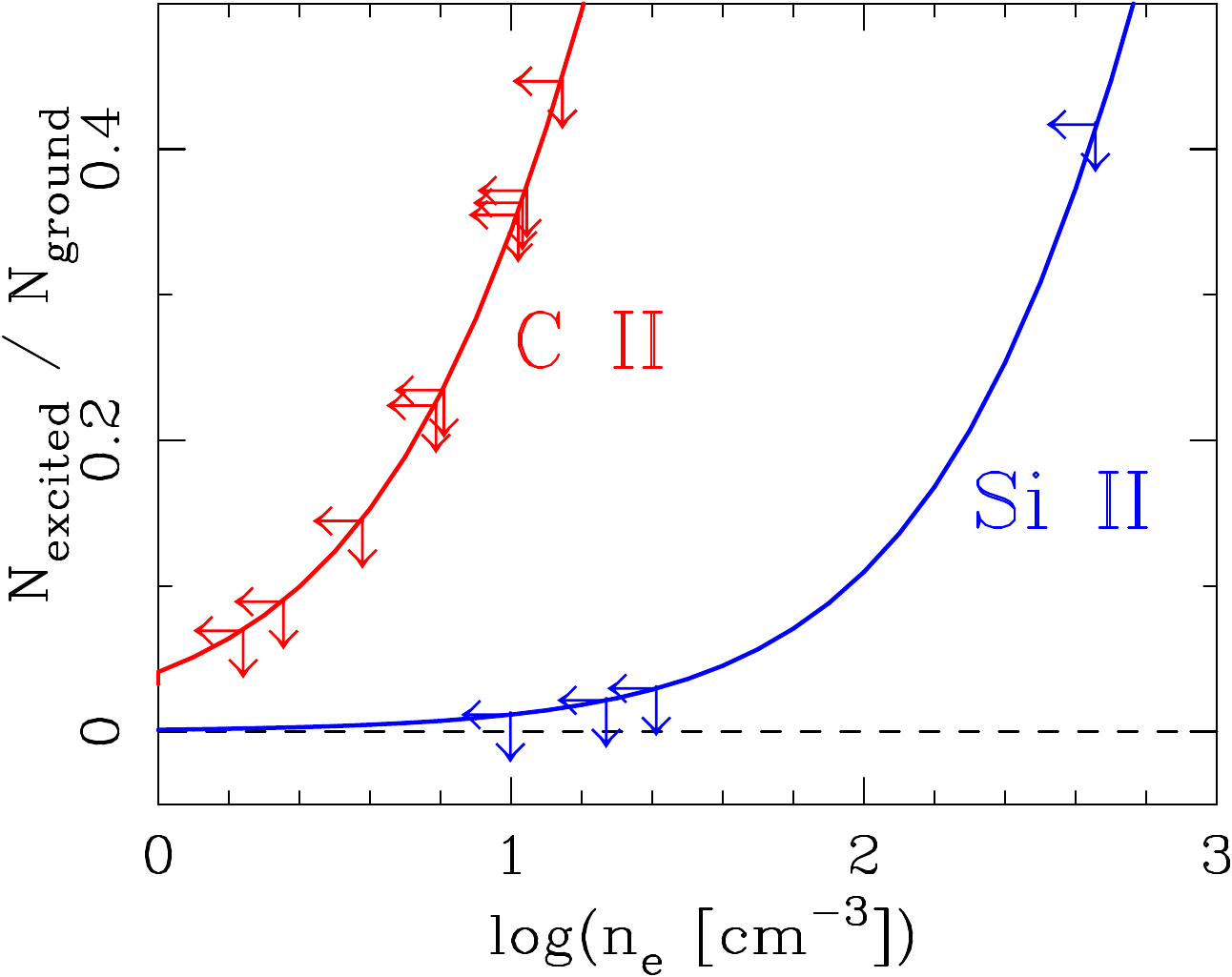}
  \end{center}  
  \caption{Column density ratios between excited and ground states of
    \ion{C}{2} and \ion{Si}{2} (red and blue curves) as a function of
    electron density, which is calculated with the {\sc CHIANTI} 10.0
    Database \citep{der97,del21} assuming a gas temperature of
    10,000~K. Arrows denote upper limits on the column density ratio
    of \ion{C}{2}$^*$ to \ion{C}{2} and \ion{Si}{2}$^*$ to \ion{Si}{2}
    and the corresponding upper limits on electron
    density.\label{fig:eden}}
\end{figure}

\begin{figure}[ht!]
  \begin{center}
    \includegraphics[width=9cm,angle=270]{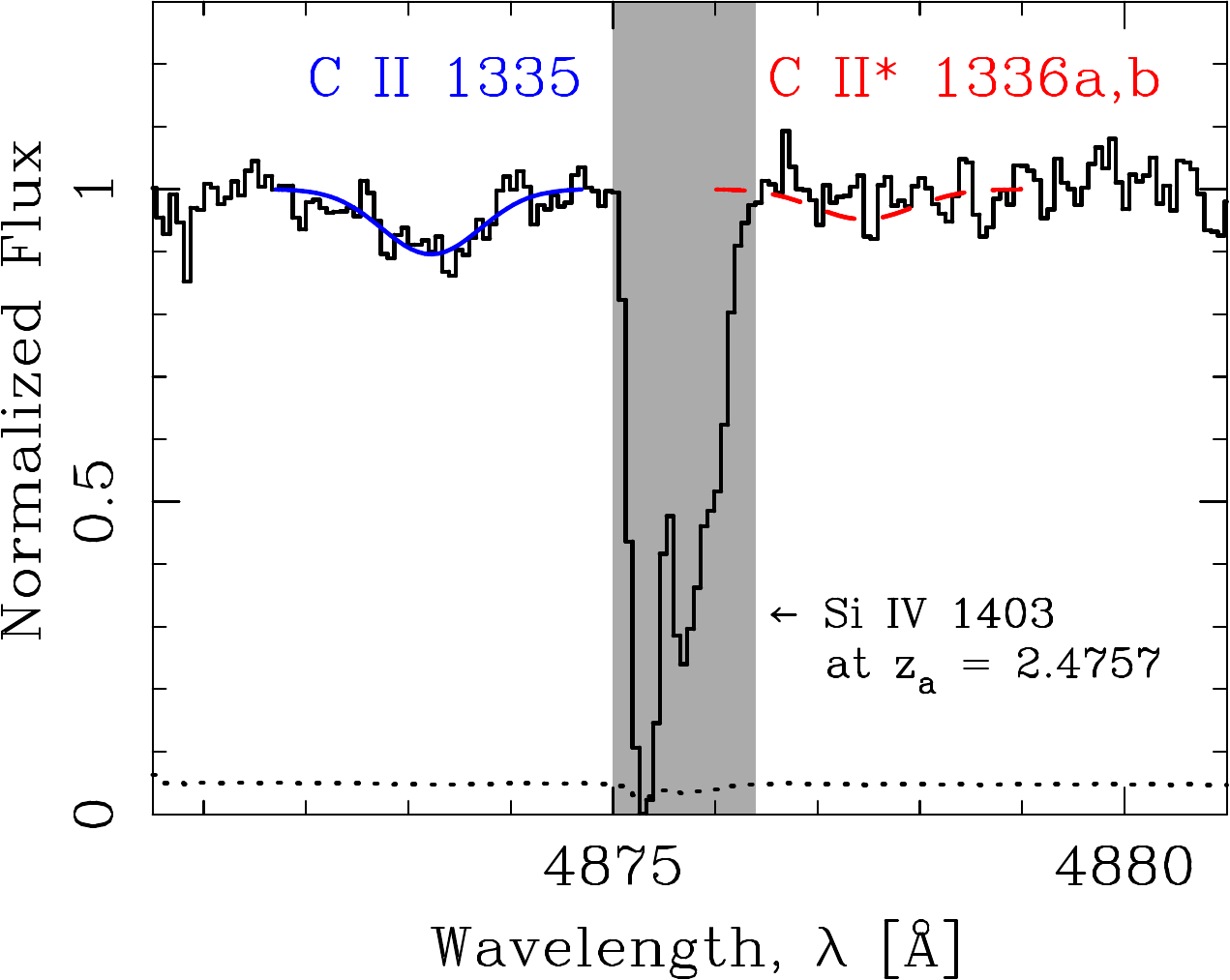}
  \end{center}  
  \caption{Normalized flux spectrum (solid histogram) and error
    spectrum (dotted histogram) around the absorption lines of
    \ion{C}{2}~1335 and \ion{C}{2}$^*$~1336a,b at \zabs\ = 2.6517 in
    Q0805+0441. The former is detected at a 3$\sigma$ level, while the
    latter is not detected.  The blue solid curve is a fitted model to
    \ion{C}{2}~1335. The red dashed curve is a 1-$\sigma$ upper limit
    on \ion{C}{2}$^*$~1336a,b. The strong absorption profile in the
    gray shaded area is an unrelated \ion{Si}{4}~1403 line at \zabs\ =
    2.4757.\label{fig:spec}}
\end{figure}

We place only upper limits on the electron density, which we plot in
Figure~\ref{fig:eden}, since i) no lines in excited states (i.e.,
\ion{C}{2}$^*$1336, \ion{Si}{2}$^*$1265, \ion{Si}{2}$^*$1533) are
detected while the corresponding lines in ground states are detected
at a $\geq 3\sigma$ level (see Figure~\ref{fig:spec} as an example)
and ii) the column densities of ground states are lower limits as
described above. Similarly, we place lower limits on the NAL outflow
radial distances using equation~(3) of \citet{nar04} (see also our
equation~(2), above). If both \ion{C}{2} and \ion{Si}{2} lines are
detected in a single NAL system, we prioritize the former because
\ion{C}{2} provides a stronger constraint. In addition to the
low-ionization lines, we also evaluate line parameters of neutral
Hydrogen (\ion{H}{1}) using only the \lya\ line\footnote{Although
  higher order Lyman series lines (e.g., \lyb) are covered by some
  spectra of our sample quasars, their corresponding signal-to-noise
  ratio is very low.}.

The column densities of \ion{H}{1}, \ion{C}{2}, and \ion{Si}{2} in
ground and excited states (only upper limits at 1$\sigma$ level),
electron densities, and radial distances from the quasar continuum
source of the 11 NAL absorbers are summarized in Table~\ref{tab:nals}.

\section{Results \& Discussion} \label{sec:res-dis}

Using the parameters we evaluated in \S\ref{sec:sample}, we calculate
the mass outflow rate and the kinetic luminosity following
\citet{bor12}:
\begin{equation}
\dot{M} = 4 \pi R f_{\rm c} \mu m_{\rm p} N_{\rm H} v_{\rm
  ej}, \label{eqn:mass}
\end{equation}
\begin{equation}
  \dot{E_{\rm k}} = \frac{1}{2} \dot{M} v_{\rm ej}^2 = 2 \pi R f_{\rm c}
  \mu m_{\rm p} N_{\rm H} v_{\rm ej}^3, \label{eqn:kinetic}
\end{equation}
where $f_{\rm c}$ is the global covering fraction of the NAL outflow,
$\mu$ (= 1.4) is the mean molecular weight, and $m_{\rm p}$ is the
proton mass. We use $f_{\rm c}$ = 0.5 since intrinsic NALs are
identified at least 50\%\ of optically luminous quasars
\citep{mis07a}. Since $f_{\rm c}$ was estimated based only on partial
coverage analysis, we caution that it is uncertain. The kinetic
luminosity and feedback efficiency estimated below are proportional to
$f_{\rm c}$, therefore they would carry the same fractional
uncertainty as $f_{\rm c}$.  By substituting parameters in
Table~\ref{tab:nals} into equations~(\ref{eqn:mass}) and
(\ref{eqn:kinetic}), we evaluate $\dot{M}$ and $\dot{E_{\rm k}}$ and
summarize the results in Table~\ref{tab:feedback}.

\begin{deluxetable}{ccccccc}
\tablecaption{Feedback Efficiency of NAL outflows \label{tab:feedback}}
\tablewidth{0pt}
\tablehead{
\colhead{(1)}   &
\colhead{(2)}   &
\colhead{(3)}   &
\colhead{(4)}   &
\colhead{(5)}   &
\colhead{(6)}   &
\colhead{(7)}   \\
\colhead{quasar}                      &
\colhead{\zabs}                       &
\colhead{$\log N_{\rm H}$$^a$}          &
\colhead{\vej}                        &
\colhead{$\dot{M}$$^b$}               &
\colhead{$\dot{E_{\rm k}}$$^c$}         &
\colhead{$\varepsilon_{\rm k}$$^d$}     \\
\colhead{}                            &
\colhead{}                            &
\colhead{(\cmm)}                      &
\colhead{(\kms)}                      &
\colhead{(M$_{\odot}$~yr$^{-1}$)}        &
\colhead{(erg~s$^{-1}$)}               &
\colhead{}                            
}
\startdata
Q0805+0441  & 2.6517 & 17.75 & 18171 & $>$7.9$\times 10^{1}$ & $>$8.24$\times 10^{45}$ & $>$0.043              \\
HS1103+6416 & 1.8919 & 18.17 & 29437 & $>$7.8$\times 10^{2}$ & $>$2.12$\times 10^{47}$ & $>$0.091              \\
Q1107+4847  & 2.7243 & 18.14 & 21385 & $>$6.9$\times 10^{2}$ & $>$1.00$\times 10^{47}$ & $>$0.047              \\
Q1330+0108  & 3.1148 & 17.62 & 27433 & $>$1.5$\times 10^{2}$ & $>$3.65$\times 10^{46}$ & $>$0.19               \\
Q1548+0917  & 2.6082 & 20.33 & 11480 & $>$8.4$\times 10^{4}$ & $>$3.48$\times 10^{48}$ & $>$7.8                \\
            & 2.6659 & 17.90 &  6727 & $>$9.7$\times 10^{1}$ & $>$1.39$\times 10^{45}$ & $>$3.1$\times 10^{-3}$ \\
            & 2.6998 & 18.34 &  3959 & $>$1.2$\times 10^{2}$ & $>$5.78$\times 10^{44}$ & $>$1.3$\times 10^{-3}$ \\
Q1554+3749  & 2.3777 & 18.27 & 24356 & $>$1.2$\times 10^{3}$ & $>$2.23$\times 10^{47}$ & $>$0.34               \\
HS1700+6416 & 2.4330 & 20.40 & 24195 & $>$3.1$\times 10^{5}$ & $>$5.78$\times 10^{49}$ & $>$22                 \\
HS1946+7658 & 2.8928 & 19.80 & 11942 & $>$3.5$\times 10^{4}$ & $>$1.57$\times 10^{48}$ & $>$0.34               \\
            & 3.0497 & 20.01 &    98 & $>$2.7$\times 10^{3}$ & $>$8.10$\times 10^{42}$ & $>$1.8$\times 10^{-6}$ \\
\enddata
\tablenotetext{a}{Total Hydrogen column density including both neutral
  and ionized Hydrogen.}
\tablenotetext{b}{Mass outflow rate.}
\tablenotetext{c}{Kinetic luminosity.}
\tablenotetext{d}{Feedback efficiency, defined as the ratio of the NAL
  kinetic luminosity of the outflow to the Eddington luminosity of the
  quasar, $\dot{E_{\rm k}} / L_{\rm Edd}$. The values in this column
  assume the absorber distances in Table~\ref{tab:nals}.}
\end{deluxetable}

\begin{figure}[ht!]
  \begin{center}
    \includegraphics[width=11cm,angle=0]{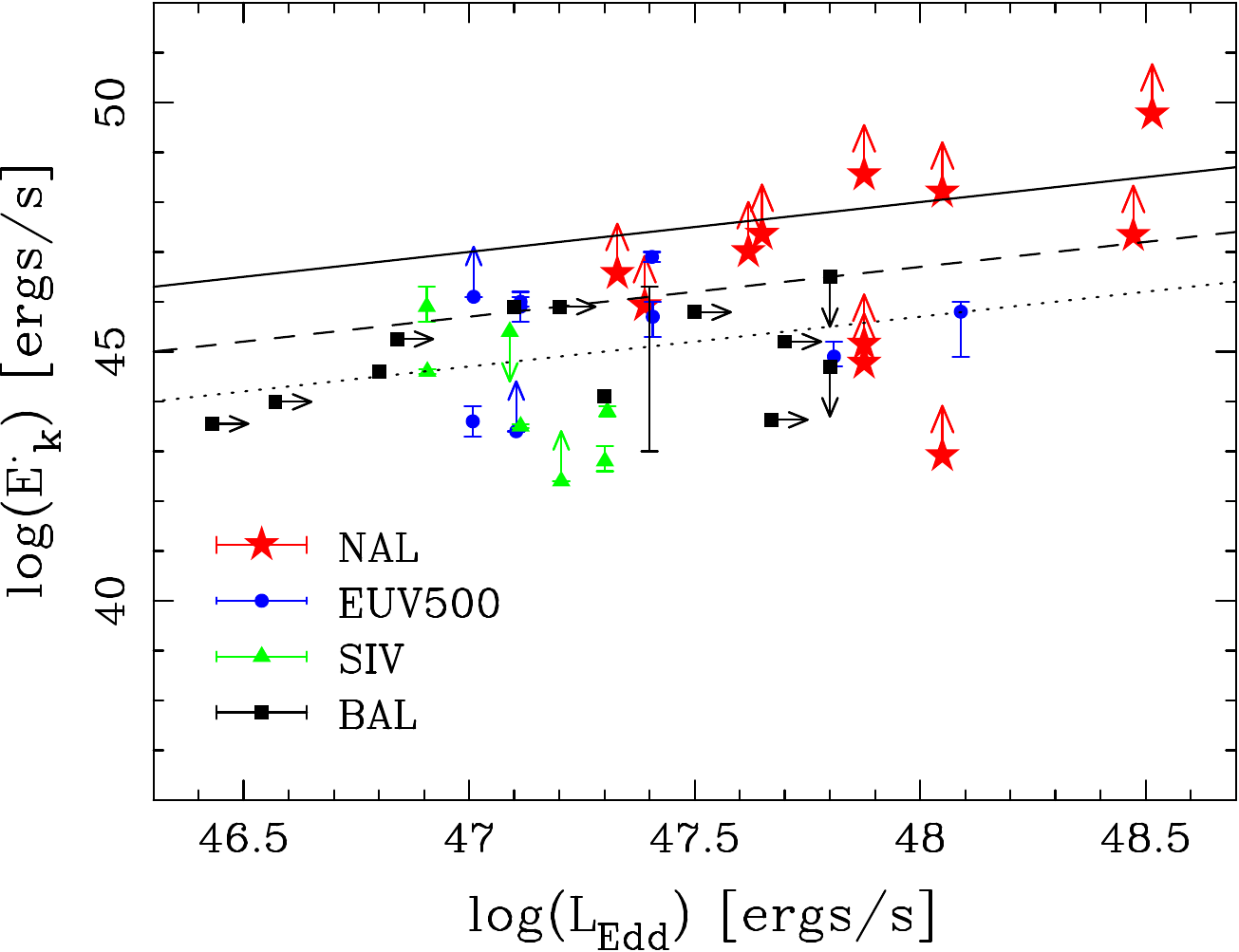}
  \end{center}  
  \caption{Kinetic luminosity ($\dot{E_{\rm k}}$) as a function of the
    Eddington luminosity ($L_{\rm Edd}$) for the 11 NAL systems of
    this study (red), along with nine EUV500 (blue) and seven
    \ion{S}{4} (green) systems from \citet{mil20}. Filled black
    squares denote BALs from the literature
    \citep{fio17,cap14,lei18,xu20,mil20,vie22}.  Upper limits (rather
    than lower limits) on $\dot{E_{\rm k}}$ are given for some
    \ion{S}{4} and BAL systems since they exhibit time variation.
    Thick, dashed, and dotted black lines are relations between
    $\dot{E_{\rm k}}$ and $L_{\rm Edd}$ for a feedback efficiency of
    1, 0.05, and 0.005, respectively.\label{fig:feedback}}
\end{figure}

The AGN feedback efficiency $\varepsilon_{\rm k}$ is computed as
\begin{equation}
  \varepsilon_{\rm k} = \frac{\dot{E_{\rm k}}}{L_{\rm Edd}} =
  \frac{\dot{E_{\rm k}}}{4 \pi G M_{\rm BH} m_{\rm p} c / \sigma_{\rm
      T}}, \label{eqn:efficiency}
\end{equation}
where $L_{\rm Edd}$ is the Eddington luminosity, $G$ is the
gravitational constant, $M_{\rm BH}$ is the mass of the SMBH, and
$\sigma_{\rm T}$ is the Thomson cross section.  The feedback
efficiencies of the 11 NAL systems in the eight quasars of this study
are summarized in Table~\ref{tab:feedback} and plotted in
Figure~\ref{fig:feedback}.

We emphasize that the mass outflow rate, $\dot{M}$, the kinetic
luminosity, $\dot{E_{\rm k}}$, and the AGN feedback efficiency,
$\varepsilon_{\rm k}$, reported above are lower limits for the
following reasons: i) equations~(\ref{eqn:mass}) and
(\ref{eqn:kinetic}) provide lower limits on $\dot{M}$ and $\dot{E_{\rm
    k}}$ if outflows are instantaneous (not continuous) with volume
filling factors of $f_{\rm V} <$ 1 (see \S5 of \citealt{bor12}) and
ii) the radial distances of the absorbers, $R$, listed in
Table~\ref{tab:nals} are lower limits since we obtain only upper
limits on the electron density $n_{\rm e}$ without detecting any
\ion{C}{2} and \ion{Si}{2} lines in excited states.

Eight of the 11 NAL systems have lower limits of the feedback
efficiency that are large enough to cause significant AGN feedback
(i.e., $\varepsilon_{\rm k}$ $>$ 0.005). Among them, the systems at
\zabs\ = 2.6082 in Q1548+0917 and at \zabs\ = 2.4330 in HS1700+6416
have extremely large feedback efficiencies greater than 7, which is
due to their relatively large Hydrogen column densities (i.e., $\log
N_{\rm H}$ $>$ 20.3). In contrast, those at \zabs\ = 2.6659 and 2.6998
in Q1548+0917 and at \zabs\ = 3.0497 in HS1946+7658 have relatively
small feedback efficiencies. The lower limits on $\varepsilon_{\rm k}$
of these systems are smaller than 0.005, which is mainly a result of
their small ejection velocities (i.e., \vej\ $<$ 7,000~\kms).
 
We also compare our result to past studies whose targets are i) BALs
\citep{fio17,cap14,lei18,xu20,mil20,vie22}, ii) highly-ionized
\ion{S}{4} outflows whose ionization potential (IP) is 47.3~eV
\citep{mil20}, and very high-ionized outflows hosting ions with IP of
more than 100~eV (e.g., \ion{Ne}{8}, \ion{Na}{9}, and \ion{Mg}{10})
whose rest-frame absorption wavelengths are in the extreme-UV region
between 500 and 1050~\AA\ (EUV500; \citealt{mil20}). As shown in
Figure~\ref{fig:feedback}, NAL outflows have an exceptionally large
efficiency compared to the other outflow classes.

Before closing the discussion, we would like to emphasize that, the
radial distances of NAL outflows we derived in \S\ref{sec:sample}
(from about 100~kpc and up to 4~Mpc) are much larger than the typical
distances of BAL and mini-BAL outflows. The distances of BAL and
mini-BAL outflows have been estimated based on i) time-variability
analysis, ii) photoionization modeling, and iii) excited/resonance
line analysis similar to that employed in this study, and are found to
range from (sub-)parsec \citep{cap14,lei18,xu20} to tens of
kilo-parsec \citep{ham01,byu22a,byu22b,wal22}.  In contrast, since
NALs are generally not observed to vary \citep{mis14} the radial
distances of NAL outflows have been constrained using only the last
two of the above methods. Thus, NAL outflows in nearby Seyfert
galaxies are found to be located at distances of several to to $\sim
10\;$kpc from the central engine
\citep[e.g.,][]{dun10,bor12}. Estimates of distances of NAL outflows
in high-$z$ quasars vary significantly.  For example, photoionization
modeling has suggested a wide range of distances from pc to several
kpc for NAL absorbers \citep[even in the same quasar;][]{wu10} while
methods similar to those used here have suggested distances greater
than $\sim 100\;$kpc \citep{ito20}. Our results here imply even larger
radial distances of that absorbing gas (i.e., several hundreds of
kpc), which place the NAL outflows away from the immediate vicinity of
the quasar central engine and in the outskirts of the host galaxy or
in the CGM \citep[cf.][]{fau12,ito20}. The small filaments that
produce the NALs may result from the interaction of a quasar-launched
blast wave with dense gas in the host galaxy, as suggested by
\citet{fau12}.  These large distances raise the question of the
connection between fast, outflowing absorbers to the quasar central
engine.  Future work should examine whether the NAL outflowing winds
indeed have traveled such a long distance from the quasar’s innermost
regions to the scale of CGM and IGM, if NALs originate in compact
parcels of gas expelled from the quasar host galaxies by quasar-driven
outflows, and if other NALs “without” low-ionized species have a
different origin (i.e., a different radial distance from the flux
source).

\section{Summary and Conclusions} \label{sec:summary}

In this study, we investigate the physical conditions of NAL outflows
to test whether they contribute to AGN feedback.  We carefully choose
11 NAL systems in eight quasars from \citet{mis07a}, and search for
excited lines of \ion{C}{2} and \ion{Si}{2} ions to place constraints
on their electron densities and radial distances from the continuum
source. Although no lines from excited states are detected, we place
lower limits on the mass outflow rates, kinetic luminosities, and
feedback efficiencies of NAL outflows for the first time. Our main
results are as follows:

\begin{itemize}

\item We study 11 NAL systems that were identified as intrinsic based
  on partial coverage analysis. We reiterate here that the selection
  method has caveats and uncertainties associated with it, which we
  discuss in \S\ref{sec:caveats}. We also note that because the
  properties of intrinsic NALs are different from those of BALs and
  mini-BALs, the corresponding absorbers are probably associated with
  a different part or phase of the quasar outflow.
 
\item The outflows traced by intrinsic NALs have large mass ejection
  rates, kinetic luminosities, and feedback efficiencies, namely $\log
  (\dot{M}/{\rm M_{\odot}~yr}^{-1}) > 79$--3.1$\times 10^{5}$, $\log
  (\dot{E_{\rm k}}/{\rm erg~s}^{-1}) > 42.9$--49.8, and
  $\varepsilon_{\rm k} \gtrsim 0$--22.

\item Eight of the 11 cases studied here have feedback efficiencies
  greater than 0.5\%\ \citep{hop10}, implying that they contribute
  significantly to AGN feedback.

\item The feedback efficiency of some NAL outflows is larger than
  those of other outflow classes including BALs, high-ionized
  \ion{S}{4}, and very high-ionized EUV500 outflows. Such high
  efficiencies suggest that the outflows traced by NALs could deliver
  significant feedback from the quasar to the host galaxy.

\item The radial distances of NAL outflows are estimated to be very
  large (i.e., hundreds of kilo-parsecs), which is larger than the
  typical distance of BAL outflows of 1~pc\,--\,10~kpc. This result
  raises the question of what is the connection of the outflows traced
  by NALs to the quasar central engine. Do intrinsic NALs trace the
  portion of the outflow that has already left the quasar host galaxy?
  The connection has to be understood before we can fully evaluate how
  the NAL gas fits into the bigger picture of quasar-driven outflows.

\item Taking these results at face value, we conclude that we need to
  take NAL outflows into account for estimating AGN feedback
  efficiency. However, additional work is needed to confirm the
  connection of NALs with partial coverage to quasar-driven outflows
  and to corroborate the results presented here.

\end{itemize}

For further investigation, we should obtain UV spectra of our sample
quasars to detect a number of metal absorption lines (both from the
ground states and excited states) in NAL systems, and place more
stringent constraints on their electron densities and radial
distances.

Our sample consists of only luminous quasars with bolometric
luminosity of $\log (L_{\rm bol}/{\rm erg~s}^{-1} ) >$ 47.2 , which
could bias the results in two ways: i) luminous quasars have larger
outflow velocity, mass outflow rate, kinetic luminosity, and feedback
efficiency \citep{gan08,fio17,bru19}, and ii) luminous quasars
over-ionize the CGM and the IGM around them both in the line-of-sight
\citep{baj88,sco00} and transverse \citep{pro13,jal19,mis22}
directions. By observing a number of fainter quasars to increase the
sample size, we will be able to study quasar feedback more generally.
Moreover, we will increase the chances of detecting \ion{C}{2} and
\ion{Si}{2} NALs in their excited states, which will help us get a
more detailed understanding of the feedback efficiency of NAL
outflows.

\newpage 

\begin{acknowledgments}
We would like to thank the anonymous referee for comments that helped
us improve the paper. We also would like to thank Christopher
Churchill for providing us with the {\sc minfit} software
packages. This work was supported by JSPS KAKENHI Grant Number 
25K01038.

The data presented herein were obtained at the W. M. Keck Observatory,
which is operated as a scientific partnership among the California
Institute of Technology, the University of California and the National
Aeronautics and Space Administration. The Observatory was made
possible by the generous financial support of the W. M. Keck
Foundation.  The authors wish to recognize and acknowledge the very
significant cultural role and reverence that the summit of MaunaKea
has always had within the indigenous Hawaiian community.  We are most
fortunate to have the opportunity to conduct observations from this
mountain.  This work was supported by JSPS KAKENHI Grant Number
21H01126.

Funding for the Sloan Digital Sky Survey IV has been provided by the
Alfred P. Sloan Foundation, the U.S.  Department of Energy Office of
Science, and the Participating Institutions.  SDSS-IV acknowledges
support and resources from the Center for High Performance Computing
at the University of Utah. The SDSS website is www.sdss.org.

SDSS-IV is managed by the Astrophysical Research Consortium for the
Participating Institutions of the SDSS Collaboration including the
Brazilian Participation Group, the Carnegie Institution for Science,
Carnegie Mellon University, Center for Astrophysics | Harvard \&
Smithsonian, the Chilean Participation Group, the French Participation
Group, Instituto de Astrof\'isica de Canarias, The Johns Hopkins
University, Kavli Institute for the Physics and Mathematics of the
Universe (IPMU) / University of Tokyo, the Korean Participation Group,
Lawrence Berkeley National Laboratory, Leibniz Institut f\"ur
Astrophysik Potsdam (AIP), Max-Planck-Institut f\"ur Astronomie (MPIA
Heidelberg), Max-Planck-Institut f\"ur Astrophysik (MPA Garching),
Max-Planck-Institut f\"ur Extraterrestrische Physik (MPE), National
Astronomical Observatories of China, New Mexico State University, New
York University, University of Notre Dame, Observat\'ario Nacional /
MCTI, The Ohio State University, Pennsylvania State University,
Shanghai Astronomical Observatory, United Kingdom Participation Group,
Universidad Nacional Aut\'onoma de M\'exico, University of Arizona,
University of Colorado Boulder, University of Oxford, University of
Portsmouth, University of Utah, University of Virginia, University of
Washington, University of Wisconsin, Vanderbilt University, and Yale
University.
\end{acknowledgments}

\end{document}